\documentclass[ALICE,manyauthors]{cernphprep}

\def\pt{\mbox{$p_{\rm T}$ }}   
\def\pbpb {Pb--Pb }
\def\ppb {p--Pb }
\def\pbp {Pb--p }
\def\pa {p--A }
\def\pp {pp }
\def\jpsi {\mbox{J/$\psi$ }}
\def\upsi {\mbox{$\Upsilon$ }}
\def\upsis {\mbox{$\Upsilon$(1S) }}
\def\upsiss {\mbox{$\Upsilon$(2S) }}

\def\rpa {\mbox{$R_{\rm{pPb}}$}}
\def\rfb {\mbox{$R_{\rm{FB}}$}}
\newcommand{\snn}  {\ensuremath{\sqrt{s_{\rm NN}}}}   

\usepackage[comma,square,numbers,sort&compress]{natbib}
\usepackage{hyperref}
\usepackage{lineno}
\begin{document}%

%%%%%%%%%%%%%%%  Title page %%%%%%%%%%%%%%%%%%%%%%%%
\begin{titlepage}
\PHyear{2014}
\PHnumber{196}      % required, will be obtained from PH
\PHdate{29 July}  % required, will be obtained from PH
%

%%% Put your own title + short title here:
\title{Production of inclusive $\Upsilon$(1S) and $\Upsilon$(2S) \\ in p--Pb collisions at $\mathbf{\sqrt{s_{{\rm NN}}} = 5.02}$~TeV}
\ShortTitle{Production of $\Upsilon$(1S) and $\Upsilon$(2S) in p--Pb collisions at $\sqrt{s_{\mathrm{NN}}} = 5.02$ TeV}   % appears on right page headers

%%% Do not change the next lines
\Collaboration{ALICE Collaboration\thanks{See Appendix~\ref{app:collab} for the list of collaboration members}}
\ShortAuthor{ALICE Collaboration} % appears on left page headers, do not change

\begin{abstract}
%!TEX root = UpsipPb13_PLB.tex
We report on the production of inclusive \upsis and \upsiss in \ppb collisions at \snn~=~5.02 TeV at the LHC. 
The measurement is performed with the ALICE detector at backward ($-4.46 < y_{{\rm cms}} < -2.96$) and forward ($2.03 < y_{{\rm cms}} < 3.53$) rapidity down to zero transverse momentum. 
The production cross sections of the \upsis and \upsiss are presented, as well as the nuclear modification factor and the ratio of the forward to backward yields of $\Upsilon$(1S). A suppression of the inclusive \upsis yield in \ppb collisions with respect to the yield from \pp collisions scaled by the number of binary nucleon-nucleon collisions is observed at forward rapidity but not at backward rapidity. 
The results are compared to theoretical model calculations including nuclear shadowing or partonic energy loss effects.

\end{abstract}
\end{titlepage}
\setcounter{page}{2}

%
%
%!TEX root = UpsipPb13_PLB.tex

%%% Introduction
\section{Introduction}

Quarkonia are bound states of a heavy quark and its anti-quark. The \jpsi family is comprised of charm and anti-charm quarks and the \upsi family of bottom and anti-bottom quarks. 
The former are commonly called charmonia and the latter bottomonia. 
In elementary pp collisions, the production of a quarkonium can be understood as the creation of a heavy-quark pair ($Q\bar{Q}$) followed by its binding into a quarkonium state with given quantum numbers~\cite{Brambilla:2010cs}. 
The first step is well described by perturbative quantum chromo-dynamics (QCD) while the second step is inherently non-perturbative. 
Three main approaches are used to describe quarkonium production in hadronic collisions: the Colour Evaporation Model (CEM)~\cite{Fritzsch:1977ay,Amundson:1996qr}, the Colour-Singlet Model (CSM)~\cite{Baier:1981uk} and the Non-Relativistic QCD (NRQCD) framework~\cite{Bodwin:1994jh}. 
However, none of those models is able to satisfactorily describe simultaneously all aspects of quarkonium production in pp collisions~\cite{Lansberg:2006dh}. 

In ultra-relativistic \pbpb collisions, quarkonia are important probes to study the properties of the deconfined state of partonic matter, the quark-gluon plasma (QGP). 
Such a state is predicted by QCD at high temperature and pressure~\cite{Collins:1974ky,Shuryak:1980tp}. 
Since quarkonia are produced at the early stage of the collision, they are expected to interact with the QGP throughout its evolution. 
In particular, in the colour-screening scenario~\cite{Matsui:1986dk} quarkonium states are suppressed in the QGP with different dissociation probabilities for the various mass states, depending on their binding energy. 
The CMS Collaboration at the Large Hadron Collider (LHC) has reported on the observation of the sequential suppression of bottomonium states in \pbpb collisions at \snn = 2.76 TeV~\cite{Chatrchyan:2012lxa,Chatrchyan:2011pe}. 
However, other hot nuclear matter effects besides colour screening, as well as cold nuclear matter (CNM) effects, do complicate this simple picture. 
On the one hand, recent measurements by the ALICE Collaboration are compatible with a regeneration mechanism playing an important role in the production of \jpsi in \pbpb collisions at the LHC~\cite{Abelev:2012rv,ALICE:2013xna,Abelev:2013ila}. 
Additional \jpsi are expected to be produced from deconfined charm quarks by kinetic recombination in the QGP~\cite{Thews:2000rj,Grandchamp:2001pf} or by statistical hadronization at the phase boundary~\cite{BraunMunzinger:2000px}.
This additional, hot nuclear matter effect, competes with the suppression by colour screening. 
Due to the lower production cross section of ${\rm b\bar{b}}$ pairs compared to ${\rm c\bar{c}}$ pairs, the regeneration of \upsis is expected to be smaller than that of J/$\psi$~\cite{Emerick:2011xu}. 
On the other hand, effects related to the presence of CNM can also modify the production of quarkonia in nucleus-nucleus collisions. 

Cold nuclear matter effects can be separated into initial and final-state effects.
Initial-state effects occur prior to the formation of the heavy-quark pair. 
These include the modification of the kinematical distribution of the partons in the nuclei compared to that in free nucleons~\cite{Eskola:2009uj,deFlorian:2003qf,deFlorian:2011fp,Hirai:2007sx} as well as parton energy loss~\cite{Gavin:1991qk,Brodsky:1992nq,Arleo:2012rs}.
First, the nuclear Parton Distribution Functions (nPDF) differ from those in free nucleons (PDF). 
Since the gluon fusion mechanism dominates the production of heavy-quark pairs in high energy collisions, quarkonium production is particularly sensitive to the gluon nPDF, which is presently not well known. 
Bjorken-$x$ ($x_{\rm Bj}$) is defined as the fraction of the hadron momentum carried by the parton.
The gluon nPDF includes a shadowing region at low $x_{\rm Bj}$ ($x_{\rm Bj} \lesssim 0.01$) corresponding to a suppression of gluons, an antishadowing region at intermediate $x_{\rm Bj}$ ($0.01 \lesssim x_{\rm Bj} \lesssim 0.3$) corresponding to an enhancement of gluons, and an additional suppression of gluons known as EMC effect at higher $x_{\rm Bj}$ ($0.35 \lesssim x_{\rm Bj} \lesssim 0.7$).
Secondly, if the quarkonium production is dominated by low $x_{\rm Bj}$ gluons, then the Colour Glass Condensate (CGC) model can be used to describe the nucleus as a coherent gluonic system that saturates at very large density~\cite{Fujii:2013gxa}.
Finally, partons can lose energy before creating the heavy-quark pair, therefore modifying the kinematic distributions of quarkonia.
Final-state effects are those that affect the heavy-quark pair during the finite time it needs to form a quarkonium state or after the state has been formed~\cite{Vogt:2010aa}. 
The $Q\bar{Q}$ pair can interact with the nuclear matter and eventually break up. 
The break-up cross section depends on the nature of the pre-resonant state and is expected to be small for $\Upsilon$(1S) at high energy~\cite{Ferreiro:2011xy,Vogt:2010aa,Lourenco:2008sk}.
The final-state resonance can also interact with surrounding comovers and lose energy or even break up~\cite{Gavin:1996yd,Capella:1996va,Ferreiro:2014bia}.
Finally, in a recent approach to parton energy loss~\cite{Arleo:2012rs}, it is hypothesized that the parton energy loss is coherent and cannot be factorized into initial and final state effects.

Cold nuclear matter effects can be studied in proton-nucleus (p--A) collisions, where the QGP is not expected to be formed. 
Charmonium states have been extensively measured in \pa collisions at various collision energies up to LHC energies. 
Bottomonium production has recently been studied thanks to the increased energy and luminosity available in collider experiments at RHIC~\cite{Adare:2012bv,Adamczyk:2013poh} and the LHC~\cite{Chatrchyan:2013nza}.
Due to the larger mass of the bottomonium states compared to the charmonium ones, the measurement of \upsi production in proton-nucleus collisions allows a study of cold nuclear matter effects in a different kinematic regime, therefore complementing the J/$\psi$ studies~\cite{Abelev:2013yxa,Aaij:2013zxa}. 
In addition, the recent measurement by the ALICE Collaboration in \pbpb collisions of a stronger \upsis suppression at forward rapidity~\cite{Abelev:2014nua} than at mid-rapidity has stressed the importance of understanding CNM effects on \upsi production (since in the colour screening scenario such a behaviour is not expected as the energy density should be larger or equal at mid-rapidity than at forward rapidity).

In this Letter, we report ALICE results on inclusive \upsi production in \ppb collisions at $\snn = 5.02$ TeV, measured via the $\mu^{+}\mu^{-}$ decay channel. 
The ALICE measurement of the $\Upsilon$(1S) and $\Upsilon$(2S) production cross section in \ppb collisions at LHC energies is presented at  backward ($-4.46 < y_{{\rm cms}} < -2.96$) and forward ($2.03 < y_{{\rm cms}} < 3.53$) centre-of-mass rapidities.
The positive rapidity is defined by the direction of the proton beam.
The \upsis production cross sections in \ppb collisions are compared to those in \pp collisions scaled by the Pb-nucleus atomic mass number $A_{\rm{Pb}} = 208$. 
This nuclear modification factor is presented as a function of rapidity. The ratio of the forward to backward yields is also discussed.

%%% Experimental apparatus and data sample
\section{Experimental apparatus and data sample}
\label{section:Apparatus}

The ALICE detector design and performance are extensively described in~\cite{Aamodt:2008zz} and~\cite{Abelev:2014ffa}. 
The analysis presented here is based on the detection of muons in the ALICE forward muon spectrometer, which covers the laboratory pseudorapidity range $-4<\eta_{\rm{lab}}<-2.5$. 
In addition, the Silicon Pixel Detector (SPD) is used to reconstruct the primary vertex, the VZERO detector provides a minimum bias trigger and the VZERO and TZERO detectors are both used as luminometers. 
A short description of these detectors is given in the following paragraphs.

The muon spectrometer consists of a set of absorbers, a  dipole magnet with a 3~Tm field integral, five tracking stations and two trigger stations. 
The front absorber, made of carbon, concrete and steel and placed between 0.9 and 5 m from the Interaction Point (IP), filters out hadrons, thus decreasing the occupancy in the tracking system. 
Muon tracking is performed by five stations, each one consisting of two planes of Cathode Pad Chambers (CPC). 
The first two stations are located upstream of the dipole magnet, the third one is embedded inside the magnet gap and the fourth and fifth are placed downstream of the dipole, just before a 1.2 m thick iron wall (7.2 interaction lengths), which absorbs secondary hadrons escaping the front absorber and low-momentum muons (having $p<1.5$~GeV/$c$ at the exit of the front absorber). 
The muon trigger system is located downstream of the iron wall and consists of two stations, each one equipped with two planes of Resistive Plate Chambers (RPC). 
The time resolution is of the order of 2~ns and the efficiency is better than 95\%~\cite{mtrBossu2012}. 
The muon trigger system delivers single muon and dimuon triggers with a programmable transverse momentum ($p_{{\rm T}}$) threshold. 
Throughout its entire length, a conical absorber around the beam pipe ($\theta<2^{\circ}$) made of tungsten, lead and steel shields the muon spectrometer against secondary particles produced by the interaction of large-$\eta$ primary particles in the beam pipe. 

Primary vertex reconstruction is performed using the SPD, the two innermost layers of the Inner Tracking System~\cite{Aamodt:2010aa}. 
It covers the pseudo-rapidity ranges $|\eta_{\rm{lab}}|<2$ and $|\eta_{\rm{lab}}|<1.4$, for the inner and outer layers, respectively. 

The two VZERO hodoscopes~\cite{Abbas:2013taa}, with 32 scintillator tiles each, are placed on each side of the IP, covering the pseudo-rapidity ranges $2.8<\eta_{\rm{lab}}< 5.1$ and $-3.7<\eta_{\rm{lab}}<-1.7$. 
Each hodoscope is segmented into 8 sectors of equal azimuthal coverage and four equal pseudo-rapidity rings. 
The logical AND of the signals from the two hodoscopes forms the Minimum Bias (MB) trigger, also used as a luminosity signal.  
A second luminosity signal is defined as the logical AND of the two TZERO arrays, located on opposite sides of the IP (\hbox{$4.6<\eta_{\rm{lab}}< 4.9$} and \hbox{$-3.3<\eta_{\rm{lab}}<-3.0$}). 
Each array consists of 12 quartz Cherenkov counters, read by photomultiplier tubes.

The data samples used for this analysis were collected in 2013. 
The number of bunches colliding at the ALICE IP ranged from 72 to 288.  
The peak luminosity during data taking was about $10^{29}$~s$^{-1}$cm$^{-2}$. 
The average number of visible interactions per bunch crossing in such conditions is about~0.06, corresponding to a multiple interaction (pile-up) probability of about 3\%.

The trigger condition used for data taking is a dimuon-MB trigger formed by the logical AND of the MB trigger and an unlike-sign dimuon trigger with a trigger probability for each of the two muon candidates that increases with \pt and is 50\% at 0.5~GeV/$c$. 
In an additional offline selection, the timing information of the two VZERO arrays is used to reject beam-halo and beam-gas events. 
The Zero Degree Calorimeters (ZDC), positioned symmetrically at 112.5~m from the IP, are used offline to reject events with a displaced vertex, originating from the interactions of satellite proton and lead bunches, as described in~\cite{Abelev:2014ffa}.

The two LHC beams  have the same magnetic rigidity but different projectile charge to mass ratio, which results in the two beams having different energies: $E_{\rm{p}}$~=~4~TeV and $E_{\rm{Pb}}/A_{\rm{Pb}} = 1.58$~TeV. 
As a consequence, the centre-of-mass system of nucleon--nucleon collisions is shifted in rapidity by $\Delta y$~=~0.465 with respect to the laboratory frame in the direction of the proton beam. 
In terms of the rapidity in the centre-of-mass frame $y_{\rm{cms}}$, the muon spectrometer acceptance is  \hbox{$2.03<y_{\rm{cms}}< 3.53$} when the proton beam is travelling in the direction of the spectrometer (p--Pb configuration), and \hbox{$-4.46<y_{\rm{cms}}<-2.96$} in the opposite case (Pb--p configuration). 
To access both rapidity ranges, data were taken in the two configurations.

About 9.3$\times$10$^6$ (2.1$\times$10$^7$)~dimuon-MB-triggered events were analyzed for the p--Pb (Pb--p) configuration, corresponding to an integrated luminosity $\mathcal{L}_{int} = 5.01 \pm 0.19$~nb$^{-1}$ ($5.81 \pm 0.20$~nb$^{-1}$). The determination of the integrated luminosities and associated uncertainties is described later.

%%% Analysis
\section{Data analysis}\label{section:analysis}

Muon track candidates are reconstructed in the muon spectrometer using the standard tracking algorithm~\cite{Aamodt:2011gj}. 
The tracks are required to exit the front absorber at a radial distance from the beam axis, $R_{\rm abs}$, in the range $17.6<R_{\rm abs}<89.5$ cm to reject tracks crossing the region of the absorber with the highest density material. 
In this region, multiple scattering and energy loss effects are large and can affect the mass resolution. 
The contribution from fake and beam-gas interaction induced tracks is reduced by selecting tracks pointing to the interaction vertex. 
In addition, tracks in the tracking system are requested to match a track segment in the trigger system (trigger tracklet).

\begin{figure}[tbh!f]
\begin{center}
\includegraphics[width=0.49\linewidth]{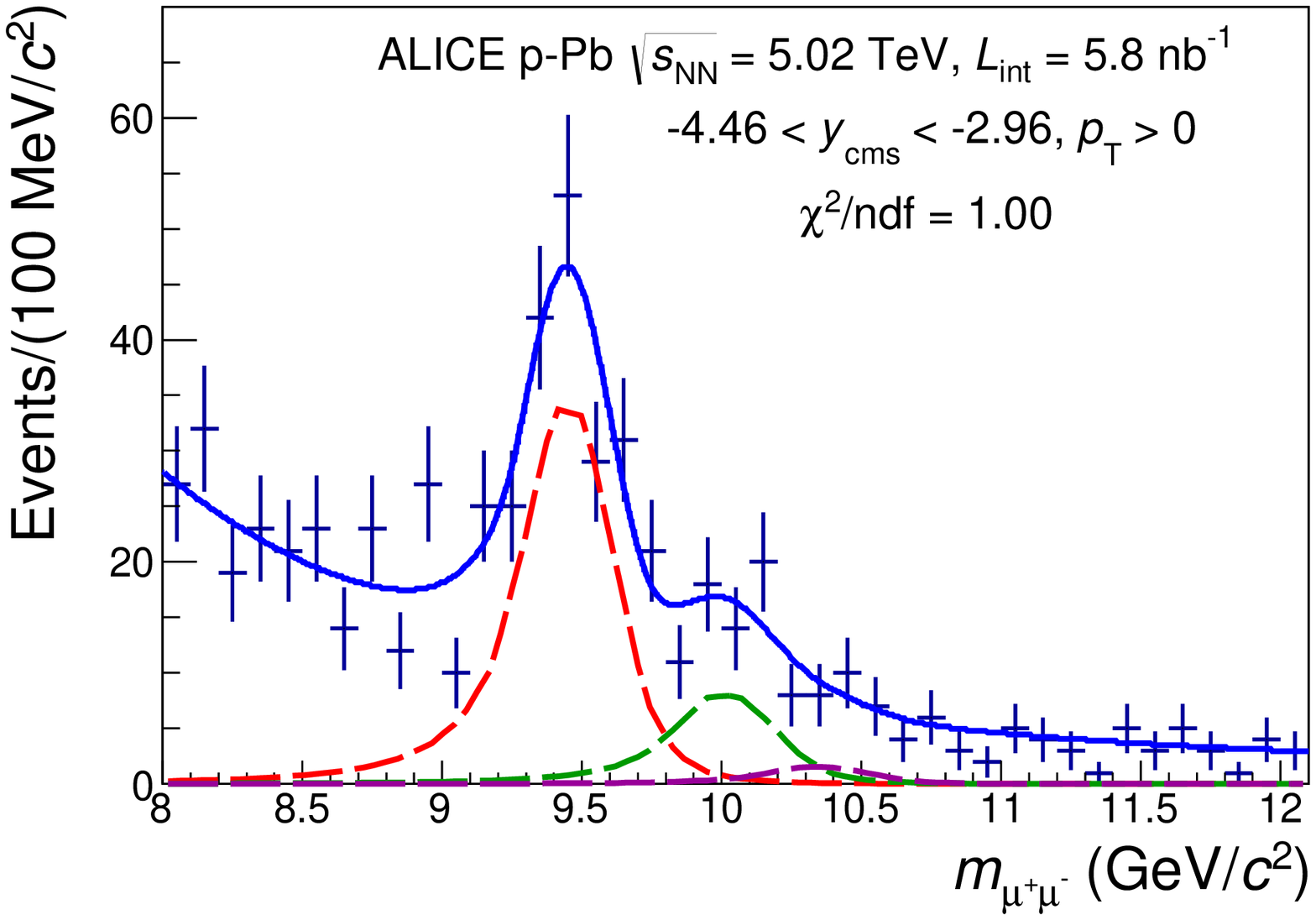} 
\includegraphics[width=0.49\linewidth]{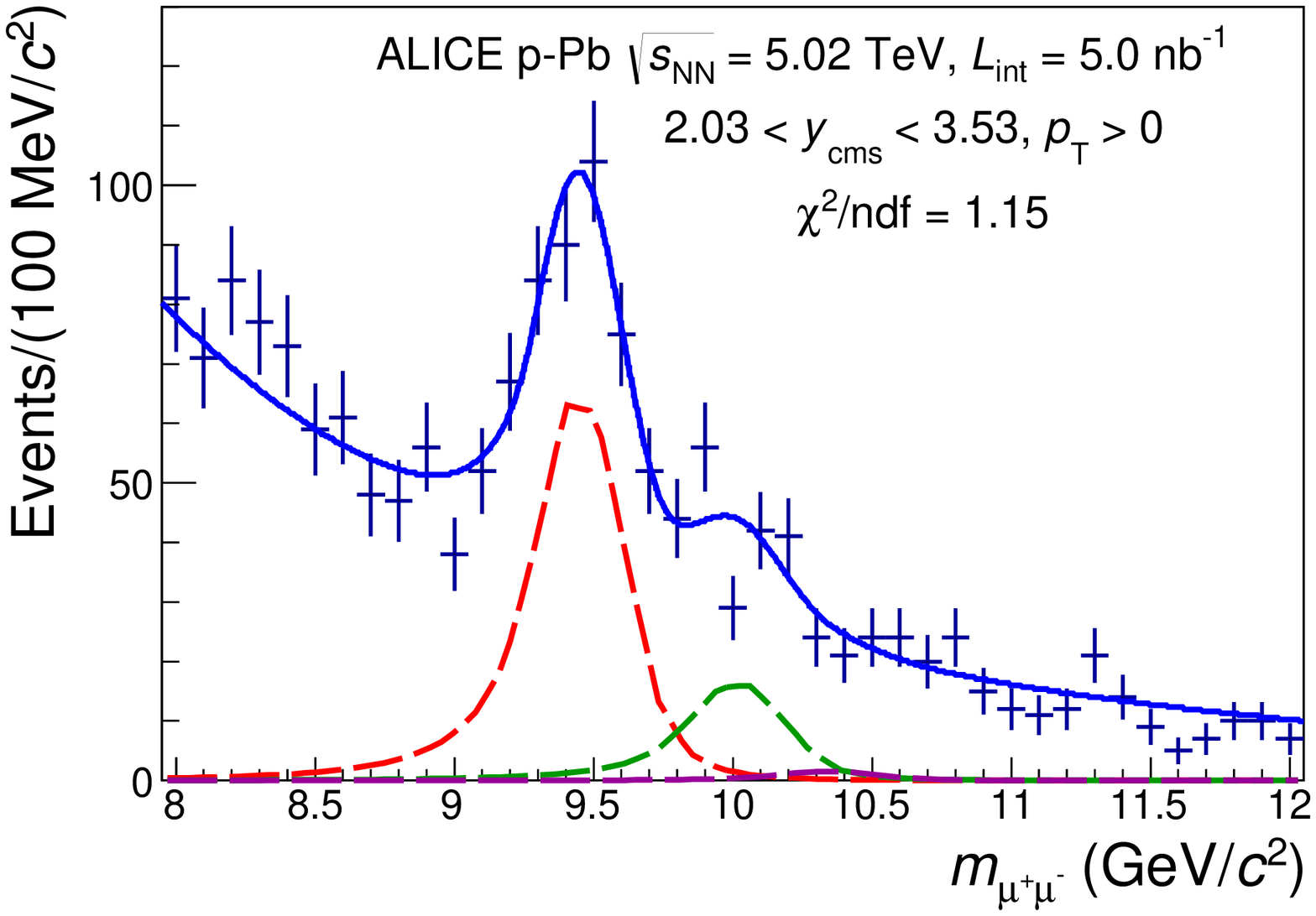} 
\end{center}
\caption{\label{fig:mass} Invariant mass distribution of opposite-sign dimuons in the rapidity regions $-4.46 < y_{\rm cms} < -2.96$ (left) and $2.03 < y_{\rm cms} < 3.53$ (right) in \ppb\ collisions. In each case, the full curve shows the total fit function and the dashed curves the signal component for the three \upsi states (see text for details).}
\end{figure}

The \upsi signal is obtained from the invariant mass distributions of opposite-sign dimuons with a laboratory pair-rapidity in the range $2.5<|y_{\rm lab}|<4$ down to zero transverse momentum. 
The raw number of $\Upsilon$ is obtained by fitting the invariant mass distributions. 
A sum of two exponential functions is used to parameterize the background continuum, and each $\Upsilon$ resonance shape is described by an extended Crystal Ball (CB) function~\cite{CB2}. 
The CB function is made of a Gaussian core and a power-law tail on each side and is found to reproduce the shape of the $\Upsilon$ peak obtained in Monte Carlo (MC) simulations. 
Since the CB tails are poorly constrained by the data, they are fixed from the results of the MC simulations. 
It is also necessary to fix the mass difference between states by using the PDG values~\cite{Beringer:1900zz} and to force the width of the $\Upsilon$(2S) and $\Upsilon$(3S) to scale proportionally with the $\Upsilon$(1S) width according to the ratio of the resonance masses. 
MC simulations validated these assumptions. 
The $\Upsilon$(1S) signal to background ratio (S/B)\footnote{The signal to background ratio and significance numbers are always evaluated determining the number of signal and background counts in an invariant mass range centred on the $\Upsilon$ mass and corresponding to $\pm 3$ times the width of the peak.} is between 0.8 to 1.8, allowing the position and width of the \upsis peak to be free parameters in the fit. 
The significance (S/$\sqrt{{\mathrm S}+{\mathrm B}}$) for $\Upsilon$(1S) is between $6.3$ and $11.6$ for the rapidity bins considered in the analysis. 
The significance of the \upsiss in the rapidity ranges $-4.46 < y_{\rm cms} < -2.96$ and $2.03 < y_{\rm cms} < 3.53$ is larger than $3$, which allows a reliable measurement. 
However, due to the limited statistics, the significance of the $\Upsilon$(3S) state is too low to separate the signal from the underlying background.
Figure~\ref{fig:mass} illustrates the fitting method for the rapidity intervals $-4.46 < y_{\rm cms} < -2.96$ (left panel) and $2.03 < y_{\rm cms} < 3.53$ (right panel). 
The measured $\Upsilon$(1S) peak position is in agreement with the resonance mass value from PDG~\cite{Beringer:1900zz} and the measured width ($155 \pm 25~{\rm MeV/}c^{2}$ in $-4.46 < y_{\rm cms} < -2.96$ and $160 \pm 22~{\rm MeV/}c^{2}$ in $2.03 < y_{\rm cms} < 3.53$) agrees with the results from MC simulations. 
A similar agreement was observed for all rapidity bins considered in this Letter. 

To investigate the systematic uncertainties on the signal extraction procedure, different fits were performed parameterizing the background continuum with the sum of two power-law functions and using alternative invariant mass fitting ranges.
Since some parameters are fixed in the fitting procedure, the related systematic uncertainties were also studied. 
The CB tail parameters were varied according to their spread obtained by several fits of the MC distributions in different mass ranges.
The width of the $\Upsilon$(2S) and $\Upsilon$(3S) were varied according to the size of the uncertainties of the $\Upsilon$(1S) width obtained from the fit. 
The latter method was similarly used to estimate the systematic uncertainty related to the fixing of the $\Upsilon$(2S) and $\Upsilon$(3S) peak position. 
The raw number of $\Upsilon$(1S) and $\Upsilon$(2S) in the rapidity range $-4.46 < y_{\rm cms} < -2.96$ are $161 \pm 21{\rm (stat)} \pm 9{\rm (syst)}$ and $42 \pm 14{\rm (stat)} \pm 5{\rm (syst)}$, respectively. 
In the $2.03 < y_{\rm cms} < 3.53$ rapidity range, they are $305 \pm 34{\rm (stat)} \pm 13 {\rm (syst)}$ for $\Upsilon$(1S) and $83 \pm 23{\rm (stat)} \pm 10{\rm (syst)}$ for $\Upsilon$(2S). 

The acceptance-times-efficiency of the muon spectrometer for the measurement of $\Upsilon$, $A\times\varepsilon$, is calculated with MC simulations. 
The $p_{\rm T}$ and $y$ distributions of the generated $\Upsilon$(1S) were extrapolated, with a procedure equivalent to the one adopted for the  \jpsi~\cite{Bossu:2011qe}, to \snn~=~5.02 TeV from existing pp measurements~\cite{Acosta:2001gv,LHCb:2012aa,Khachatryan:2010zg}. 
Nuclear shadowing calculations~\cite{Eskola:1998df} were used to include the expected CNM effects. 
The systematic uncertainty was estimated by varying the $p_{\rm T}$ and $y$ input distributions by an amount sufficiently large (based on theoretical estimations) to include the a priori unknown impact of CNM effects. 
Since the available data favour a zero or small polarization of $\Upsilon$(1S)~\cite{Abazov:2008aa,Russ:2012tp,Chatrchyan:2012woa}, an unpolarized production was assumed. 
Particle transport is performed using GEANT3~\cite{Brun:1994aa} and a realistic detector response is applied to the simulated hits in order to reproduce the performance of the apparatus during data taking. 
The time dependence of the tracking and trigger efficiencies is taken into account by incorporating in the MC simulations the dead channel maps obtained from the online detector information and the trigger chamber efficiencies obtained from a real data analysis. 
In addition, a realistic description of the residual misalignment of the tracking chambers is included in the simulations. 
The tracking efficiency is evaluated with data by analyzing the cluster distribution of the reconstructed tracks in the detection chambers with the algorithm described in~\cite{Aamodt:2011gj}. 
The same algorithm can be used to estimate the tracking efficiency from MC data. 
The systematic uncertainties on this value are obtained by comparing the tracking efficiency estimated from real and MC data. 
The efficiency of the muon triggering system is calculated from data and results from the analysis of trigger tracklet distributions reconstructed from clusters in the four planes of the two trigger stations. 
The corresponding systematic uncertainties are obtained by varying the trigger chamber efficiency in MC simulations by an amount equivalent to the statistical uncertainties on the real data estimation. 
The quality of the matching of the tracking and triggering system information is ensured by a $\chi^{2}$ cut. 
In order to quantify the systematic uncertainties on the matching efficiency, the cut was varied in the same proportions while analyzing both real and MC data. The observed difference in the matching probabilities provides the uncertainties.  

The $A\times\varepsilon$ values and the corresponding systematic uncertainties for $\Upsilon{\rm (1S)}$ measured during the \ppb and the \pbp data taking periods are $(29.0\pm2.0)\%$ and $(20.1\pm1.6)\%$, respectively. 
The value of $A\times\varepsilon$ is lower for the \pbp period mainly due to a reduced tracking efficiency. 
The $\Upsilon$(2S) $A\times\varepsilon$ and the corresponding systematic uncertainties were evaluated with the same method and the same input distributions as for the $\Upsilon$(1S). 
The observed differences between the $\Upsilon$(2S) and $\Upsilon$(1S) $A\times\varepsilon$ are less than $0.5\%$. 
The shape variations between the different input distributions used in the study of the $A\times\varepsilon$ systematic uncertainties were large enough to cover the differences between the $\Upsilon$(1S) and $\Upsilon$(2S) distributions observed by LHCb in the rapidity range $2 < y_{{\rm cms}} < 4.5$ in pp collisions~\cite{Aaij:2014nwa,LHCb:2012aa,Aaij:2013yaa}.

The raw number of $\Upsilon$(1S) obtained with the fit procedure described previously, $N[\Upsilon{\rm (1S)}]$, is corrected for the branching ratio of the dimuon decay channel, ${\rm BR}_{\Upsilon{\rm (1S)}\rightarrow\mu^{+}\mu^{-}} = 0.0248\pm0.0005$~\cite{Beringer:1900zz} and for the acceptance-times-efficiency, $(A\times\varepsilon)_{\Upsilon{\rm (1S)}}$. 
The $\Upsilon$(1S) cross section is obtained as
\begin{center}
\begin{equation}
\sigma^{\Upsilon{\rm (1S)}}_{\mathrm{pPb}}~=~\frac{N[\Upsilon{\rm (1S)}]/(A\times\varepsilon)_{\Upsilon{\rm (1S)}}}{{\rm BR}_{\Upsilon{\rm (1S)}\rightarrow\mu^{+}\mu^{-}}\times \mathcal{L}},
\label{XSectionFormula}
\end{equation}
\end{center}
where the integrated luminosity $\mathcal{L}$~=~$N_{\rm{MB}}/\sigma_{\rm{MB}}$ is the ratio between the number of MB events and the MB trigger cross section. 
Since the analyzed data sample is made of dimuon triggered events, it is necessary to use a scaling factor, $F$, to obtain the number of MB events from the number of triggered events. 
The inverse of the $F$ factor corresponds to the probability of having the dimuon trigger condition verified in an MB event. 
Its average value is $F = 1129 \pm 2{\rm (stat)} \pm 11{\rm (syst)}$ and $F = 589 \pm 2{\rm (stat)} \pm 6{\rm (syst)}$ for the \ppb and \pbp data taking periods, respectively. 
These values and the corresponding statistical uncertainties were obtained by averaging the results of two different methods, one based on the ratio of trigger rates and the other based on the offline selection of dimuon events in the MB data sample~\cite{Abelev:2013yxa}. 
The systematic uncertainties reflect the difference between the results obtained with the two methods.
The MB trigger cross section $\sigma_{\rm{MB}}$ was measured with a van der Meer scan~\cite{Vandermeer} and found to be $2.09 \pm 0.07$~b ($2.12 \pm 0.07$~b) for the p--Pb (Pb--p) configuration, where the uncertainties for the two configurations are partially correlated~\cite{Abelev:2014epa}.
The luminosity was also independently determined, in a similar way, by means of the TZERO-based luminosity signal. 
The two measurements differ by at most 1\% throughout the whole data-taking period. 
Such a small variation was combined quadratically with the $N_{\rm{MB}}$ and $\sigma_{\rm{MB}}$ uncertainties, to get a total luminosity uncertainty of 3.8\% for the p--Pb configuration (forward rapidities) and 3.5\% for  the Pb--p configuration (backward rapidities). 

%%% Results
\section{Results}\label{section:results}

The \upsis production cross sections in \ppb collisions at \snn~=~5.02 TeV are:
\begin{equation*}
\sigma^{\Upsilon{\rm (1S)}}_{\mathrm{pPb}}(-4.46 < y_{{\rm cms}} < -2.96) = 5.57 \pm 0.72 ({\rm stat}) \pm 0.60 ({\rm syst})~\mu\rm{b}, 
\end{equation*}
\begin{equation*}
\sigma^{\Upsilon{\rm (1S)}}_{\mathrm{pPb}}(2.03 < y_{{\rm cms}} < 3.53) = 8.45 \pm 0.94 ({\rm stat}) \pm 0.77 ({\rm syst})~\mu\rm{b}. 
\end{equation*} 

The $\Upsilon$(2S) production cross sections in \ppb collisions at \snn~=~5.02 TeV, obtained in a similar way but with ${\rm BR}_{\Upsilon{\rm (2S)}\rightarrow\mu^{+}\mu^{-}} = 0.0193\pm0.0017$~\cite{Beringer:1900zz}, are:
\begin{equation*}
\sigma^{\Upsilon{\rm (2S)}}_{\mathrm{pPb}}(-4.46 < y_{{\rm cms}} < -2.96) = 1.85 \pm  0.61({\rm stat}) \pm 0.32 ({\rm syst})~\mu\rm{b}, 
\end{equation*}
\begin{equation*}
\sigma^{\Upsilon{\rm (2S)}}_{\mathrm{pPb}}(2.03 < y_{{\rm cms}} < 3.53) = 2.97 \pm 0.82 ({\rm stat}) \pm 0.50 ({\rm syst})~\mu\rm{b}. 
\end{equation*}

A summary of the different sources of systematic uncertainties and their relative value is given in Table~\ref{tab:table1}. 
The uncertainties of type II are not fully uncorrelated with rapidity and no trivial factorization in correlated and uncorrelated parts can be made. Hence, they are labelled as uncorrelated, but they cannot be quadratically combined to obtain the rapidity integrated result.

\begin{table}[tbh!f]
\begin{center}
\begin{tabular}{lcc}
\textrm{Source}&
\multicolumn{1}{	c}{\textrm{Backward rapidity}} & \textrm{Forward rapidity}\\
\hline
\hline
\textrm{Signal extraction: $\Upsilon$(1S)} & 5\%-6\%(II) & 4\%-6\%(II) \\
\textrm{Signal extraction: $\Upsilon$(2S)} & 12\%(II) & 12\%(II) \\
\hline
\textrm{Input MC parameterization: $\Upsilon$(1S)} &  2\%-5\%(II) & 4\%-6\%(II)\\
\textrm{Input MC parameterization: $\Upsilon$(2S)} &  5\%(II) & 5\%(II)\\
\textrm{Tracking efficiency}& 6\%(II) & 4\%(II)\\
\textrm{Trigger efficiency}& 2\%(II) & 2\%(II)\\
\textrm{Matching efficiency}& 1\%(II)& 1\%(II) \\
\hline
$\sigma_{\rm pp}^{\Upsilon{\mathrm (1S)}}$ (interpolation) &  11\%-13\% (II) & 7\%-12\%(II)\\
$\mathcal{L}$ (correlated)&   1.6\%(I) & 1.6\%(I)\\
$\mathcal{L}$ (uncorrelated)&      3.1\%(II) & 3.4\%(II)\\
\end{tabular}
\caption{\label{tab:table1} Summary of the relative systematic uncertainties on each quantity entering in the calculations of the results. Type I (II) stands for uncertainties correlated (uncorrelated) with rapidity. Type II uncertainties are given as a range including the smallest and the largest values observed in the bins considered in this analysis. Results are presented for the backward ($-4.46 < y_{\rm cms} < -2.96$) and forward ($2.03 < y_{\rm cms} < 3.53$) rapidity regions.}
\end{center}
\end{table}

\begin{figure}[tbh!f]
\begin{center}
\includegraphics[width=0.75\linewidth]{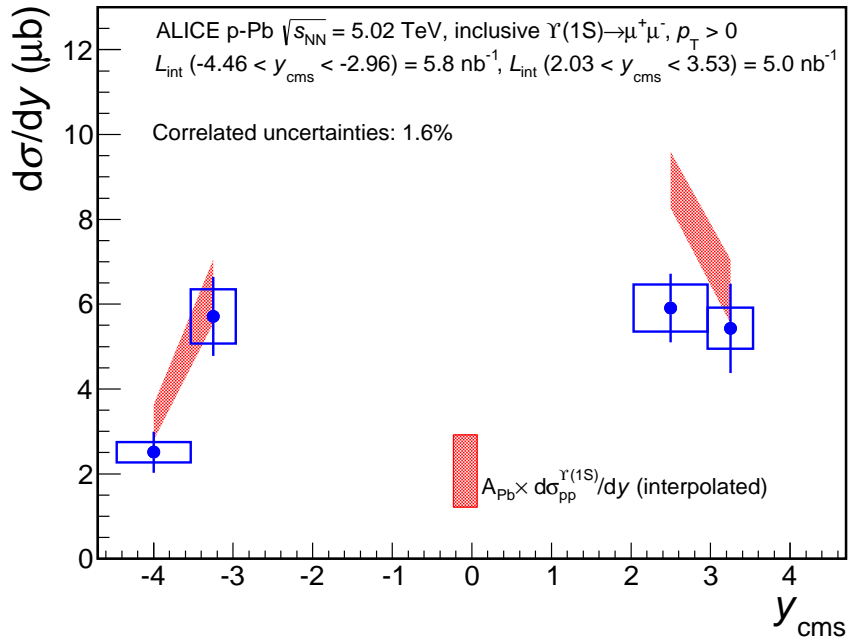} 
\end{center}
\caption{\label{fig:xsec} Inclusive \upsis production cross section as a function of rapidity in \ppb collisions at \snn~=~5.02 TeV. 
The vertical error bars represent the statistical uncertainties and the open boxes the uncorrelated systematic uncertainties. 
The correlated systematic uncertainty is 1.6\% and is directly quoted in the figure. It is obtained by summing in quadrature the correlated uncertainty on the integrated luminosity and the uncertainty on the branching ratio of $\Upsilon$(1S) to dimuon.
The bands correspond to the inclusive \upsis pp cross section obtained with the procedure described in the text and scaled by $A_{\mathrm{Pb}}$.
}
\end{figure}

The \upsis candidates were further divided in four rapidity ranges, namely $-4.46 < y_{{\rm cms}} < -3.53$, $-3.53 < y_{{\rm cms}} < -2.96$, $2.03 < y_{{\rm cms}} < 2.96$ and $2.96 < y_{{\rm cms}} < 3.53$. 
Two of them are symmetric with respect to $y_{\rm cms}=0$.
Figure~\ref{fig:xsec} shows the inclusive \upsis differential cross section d$\sigma$/d$y$ as a function of rapidity. 
The vertical error bars represent the statistical uncertainties and the open boxes the uncorrelated systematic uncertainties. 
Also shown is the inclusive \upsis $y$-differential interpolated cross section in pp collisions at the same centre-of-mass energy (obtained as explained later in the text) scaled by $A_{\mathrm{Pb}}$. 

The CNM effects can be quantified with the nuclear modification factor,
\begin{center}
\begin{equation}
R_{\rm pPb}^{\Upsilon{\rm (1S)}}=\frac{\sigma^{\Upsilon{\rm (1S)}}_{\mathrm{pPb}}}{A_{\mathrm{Pb}}\times\sigma_{\rm pp}^{\Upsilon{\rm (1S)}}},
\label{NuclModFactFormula}
\end{equation}
\end{center}

\noindent where $\sigma_{\rm pp}^{\Upsilon{\rm (1S)}}$ is the \upsis cross section in pp collisions at $\sqrt{s}$~=~5.02 TeV. 

Since $\sigma_{\rm pp}^{\Upsilon{\rm (1S)}}$ at $\sqrt{s}$~=~5.02 TeV has not yet been measured, it was computed using a data driven $\sqrt{s}$ interpolation method.
A detailed description of the adopted procedure is given in~\cite{ppRef}. 
The LHCb Collaboration has measured the $\Upsilon{\rm(1S)}$ cross section in pp collisions at $\sqrt{s}$~=~2.76, 7 and 8~TeV, over the ranges $p_{\rm{T}}$~$<$~15~GeV/$c$ and 2~$<$~$y$~$<$~4.5, in 5 rapidity bins of equal size~\cite{Aaij:2014nwa,LHCb:2012aa,Aaij:2013yaa}. 
The LHCb results were re-binned to obtain the cross section in (approximately) the rapidity ranges of interest for this analysis: 2~$<$~$y$~$<$~3, 2~$<$~$y$~$<$~3.5, 3~$<$~$y$~$<$~3.5, 3~$<$~$y$~$<$~4.5, and 3.5~$<$~$y$~$<$~4.5. 
For each bin, the cross section as a function of energy was fitted according to 21 different shapes: 15 are based on Leading Order CEM (LO-CEM) calculations for $\Upsilon$ production, corresponding to various choices of PDFs and of the factorization scale; 3 are based on the energy-dependence of bare bottom-quark pair production (FONLL)~\cite{fonll}; the remaining three are a power law, a linear and an exponential function. 
The obtained fit parameters were used to compute the cross section at $\sqrt{s}$~=~5.02~TeV.  In order to take into account the rather poor agreement of the data with the fitting functions ($\chi^2/ndf > 2$ for all fits, where $ndf$ is the number of degrees of freedom), all the uncertainties on the fit results were rescaled by~$\sqrt{\chi^2/ndf}$. 
Fits with $\chi^2/ndf$ values larger than three times the minimum value obtained for the rapidity range considered were discarded. 
The weighted average of the surviving results was computed (using the rescaled fit uncertainty as a weight) and retained as central value. 
The average (rescaled) fit-result uncertainty was evaluated for each rapidity bin: it ranges from 7\% to 12\%. 
As an additional uncertainty, the maximum difference between the average and the individual fit results was computed: it ranges from 2\% to 7\%. 
Finally, a third uncertainty was considered, to take into account the shift of 0.035 rapidity units between the ranges adopted in the interpolation procedure and those used for the measurement of $R_{\rm pPb}^{\Upsilon{\rm (1S)}}$. 
Such an uncertainty is quantified by the maximum difference between the cross sections in the two ranges, evaluated with the LO-CEM and FONLL models, and amounts to  1\% for the forward rapidity region and 3\% for the backward rapidity region. 
Since the interpolation is performed separately for each rapidity range, the associated uncertainties are assumed to be uncorrelated with rapidity. 
For the forward and backward rapidity ranges used for the integrated results, the obtained interpolated cross-sections times branching ratio are $1451 \pm 114 {\rm (syst)}$ pb and $770 \pm 87 {\rm (syst)}$ pb, respectively.

Using the interpolated values of $\sigma_{\rm pp}^{\Upsilon{\rm (1S)}}$,  the  nuclear modification factors are
\begin{equation*}
\rm{R}^{\Upsilon{\rm (1S)}}_{\mathrm{pPb}}(-4.46 < y_{{\rm cms}} < -2.96) 
= 0.86 \pm 0.11 ({\rm stat}) \pm 0.13 ({\rm uncorr}) \pm 0.01({\rm corr}),
\end{equation*}
\begin{equation*}
\rm{R}^{\Upsilon{\rm (1S)}}_{\mathrm{pPb}}(2.03 < y_{{\rm cms}} < 3.53) 
= 0.70 \pm 0.08 ({\rm stat}) \pm 0.08 ({\rm uncorr}) \pm 0.01({\rm corr}).
\end{equation*}

\begin{figure}[tbh!f]
\begin{center}
\includegraphics[width=0.75\linewidth]{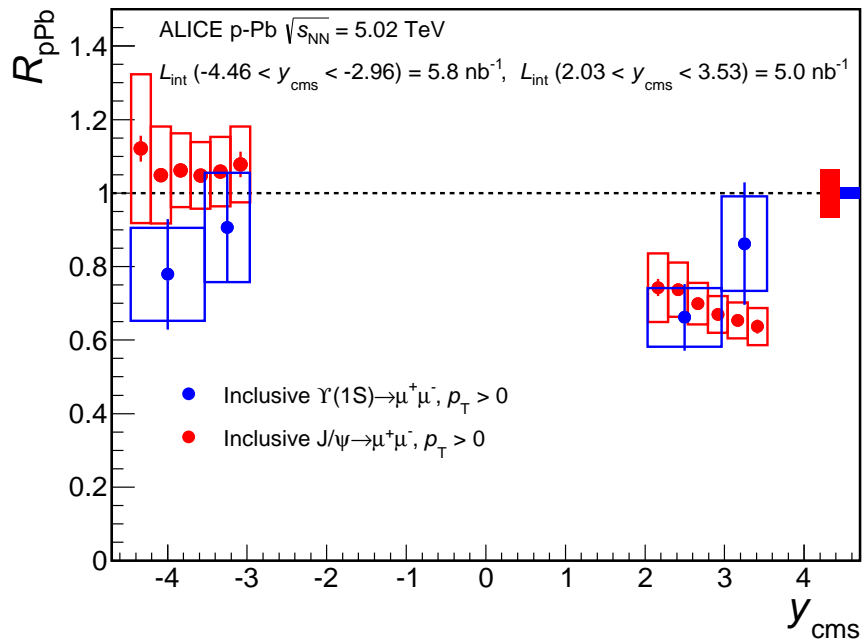} 
\end{center}
\caption{\label{fig:rppbuj} Nuclear modification factor of inclusive \upsis in \ppb collisions at \snn~=~5.02 TeV. The results are compared to those for inclusive \jpsi~\cite{Abelev:2013yxa}. The vertical error bars represent the statistical uncertainties and the open boxes the uncorrelated systematic uncertainties (for the J/$\psi$, the uncorrelated and partially correlated uncertainties have been added in quadrature). The full boxes around \rpa~=~1 show the size of the correlated uncertainties, which in the case of the $\Upsilon$ include only the correlated uncertainty on the luminosity (see Tab.~\ref{tab:table1}).}
\end{figure}

Under the assumption of a $2 \rightarrow 1$ production process ($gg \rightarrow {\Upsilon}$), the sampled $x_{\rm Bj}$ ranges are $ 5.5 \cdot 10^{-5} < x_{\rm Bj} < 2.5 \cdot 10^{-4}$ and $ 3.6 \cdot 10^{-2} < x_{\rm Bj} < 1.6 \cdot 10^{-1}$ at forward and backward rapidity, respectively.
Thus, the measurement at forward rapidity tests the shadowing region and the one at backward rapidity the anti-shadowing region.
In the case of a $2 \rightarrow 2$ production process ($gg \rightarrow {\Upsilon}g$) the covered $x_{\rm Bj}$ ranges are naturally expected to be enlarged.
In Fig.~\ref{fig:rppbuj} the inclusive \upsis nuclear modification factor in \ppb collisions at \snn = 5.02 TeV is shown in four classes of rapidity. 
The vertical error bars represent the statistical uncertainties and the open boxes the uncorrelated systematic uncertainties. 
An additional correlated uncertainty is indicated by the full box around \rpa~=~1. The \rpa\ shows a suppression of the inclusive \upsis production yields at forward rapidity in \ppb compared to pp collisions. 
At backward rapidity, the \upsis \rpa\ is compatible with unity within uncertainties, and therefore does not favour a strong gluon anti-shadowing.
Also shown in Fig.~\ref{fig:rppbuj} is the ALICE measurement of the inclusive \jpsi \rpa~\cite{Abelev:2013yxa}.
Although the uncertainties are large, it appears that at positive $y_{{\rm cms}}$ the \upsis and \jpsi \rpa\ are rather similar. 
It is worth noting that due to its larger mass, the \upsis \rpa\ at forward rapidity is higher than the \jpsi one according to all available model calculations~\cite{Albacete:2013ei,Ferreiro:2011xy,Arleo:2012rs,Fujii:2013gxa}.
At negative rapidities, the \jpsi \rpa\ are systematically above the \upsis one but the two \rpa\ are consistent within uncertainties. 
Although the rapidity ranges are not identical, the \rpa\ measured by LHCb~\cite{Aaij:2014mza} are consistent with the ALICE measurements within uncertainties, albeit systematically larger~\cite{ppRef}.

The ratio $[\Upsilon{\rm (2S)}/\Upsilon{\rm (1S)}]$ of the production cross section of $\Upsilon{\rm (2S)}\rightarrow\mu^{+}\mu^{-}$ to $\Upsilon{\rm (1S)}\rightarrow\mu^{+}\mu^{-}$ can be obtained as
\begin{equation}
[\Upsilon{\rm (2S)}/\Upsilon{\rm (1S)}] = \frac{N[\Upsilon{\rm (2S)}]/(A\times\varepsilon)_{\Upsilon{\rm (2S)}}}{N[\Upsilon{\rm (1S)}]/(A\times\varepsilon)_{\Upsilon{\rm (1S)}}}.
\label{RFormula}
\end{equation}

The branching ratio of the dimuon decay channel does not enter the calculation. 
Additionally, since the same data sample is used, $\mathcal{L}$ cancels out in the ratio. 
The systematic uncertainties on the ratios were obtained by quadratically combining the systematic uncertainties entering in each element of Eq.~\ref{RFormula}. 
Nevertheless, since the decay kinematics of the two $\Upsilon$ states are close, the systematic uncertainties on tracking, trigger and matching efficiency, estimated for the same detector in the same working conditions, cancel out in the ratio.
The results are:
\begin{equation*}
[\Upsilon{\rm (2S)}/\Upsilon{\rm (1S)}]_{\mathrm{pPb}}(-4.46 < y_{{\rm cms}} < -2.96) = 0.26 \pm 0.09 ({\rm stat}) \pm 0.04 ({\rm syst}),
\end{equation*}
\begin{equation*}
[\Upsilon{\rm (2S)}/\Upsilon{\rm (1S)}]_{\mathrm{pPb}}(2.03 < y_{{\rm cms}} < 3.53) =  0.27 \pm 0.08 ({\rm stat}) \pm 0.04({\rm syst}).
\end{equation*}

The same ratio has been measured by ALICE in pp collisions at $\sqrt{s} = 7$ TeV in the rapidity range $2.5 < y_{{\rm cms}} < 4.0$~\cite{Abelev:2014qha} and is $0.26 \pm 0.08({\rm tot})$, where the uncertainty is the quadratic sum of the statistical and systematic uncertainties. The LHCb Collaboration has measured the same ratio in pp collisions at $\sqrt{s} = 2.76$, 7 and 8 TeV and as a function of rapidity in the range $2.0 < y_{{\rm cms}} < 4.5$~\cite{Aaij:2013yaa,Aaij:2014nwa,LHCb:2012aa}. The measured $[\Upsilon{\rm (2S)}/\Upsilon{\rm (1S)}]$ is found to be, within uncertainties, independent of $\sqrt{s}$ and rapidity. For $p_{\rm T} < 15$ GeV/$c$ (14 GeV/$c$ for 8 TeV) the measured values in the range $3.0 < y_{{\rm cms}} < 3.5$ are $0.22 \pm 0.03({\rm tot})$, $0.24 \pm 0.02({\rm tot})$ and $0.25 \pm 0.01({\rm tot})$ for $\sqrt{s} = 2.76$, 7 and 8 TeV, respectively. 
Our measured ratio $[\Upsilon{\rm (2S)}/\Upsilon{\rm (1S)}]$ in \ppb collisions is compatible with the same ratio in \pp collisions. 
Within our uncertainties, there is therefore no evidence of a different magnitude of CNM effects for the \upsiss with respect to the $\Upsilon{\rm (1S)}$. 
At mid-rapidity, however, the CMS Collaboration has measured the double ratio, {\emph i.e.} the ratio $[\Upsilon{\rm (2S)}/\Upsilon{\rm (1S)}]$ in \ppb divided by that in \pp collisions, to be  $0.83 \pm 0.05({\rm stat}) \pm 0.05({\rm syst})$, suggesting a stronger suppression of the \upsiss than of the \upsis in p--Pb collisions~\cite{Chatrchyan:2013nza}.

\begin{figure}[tbh!f]
\begin{center}
\includegraphics[width=0.495\linewidth]{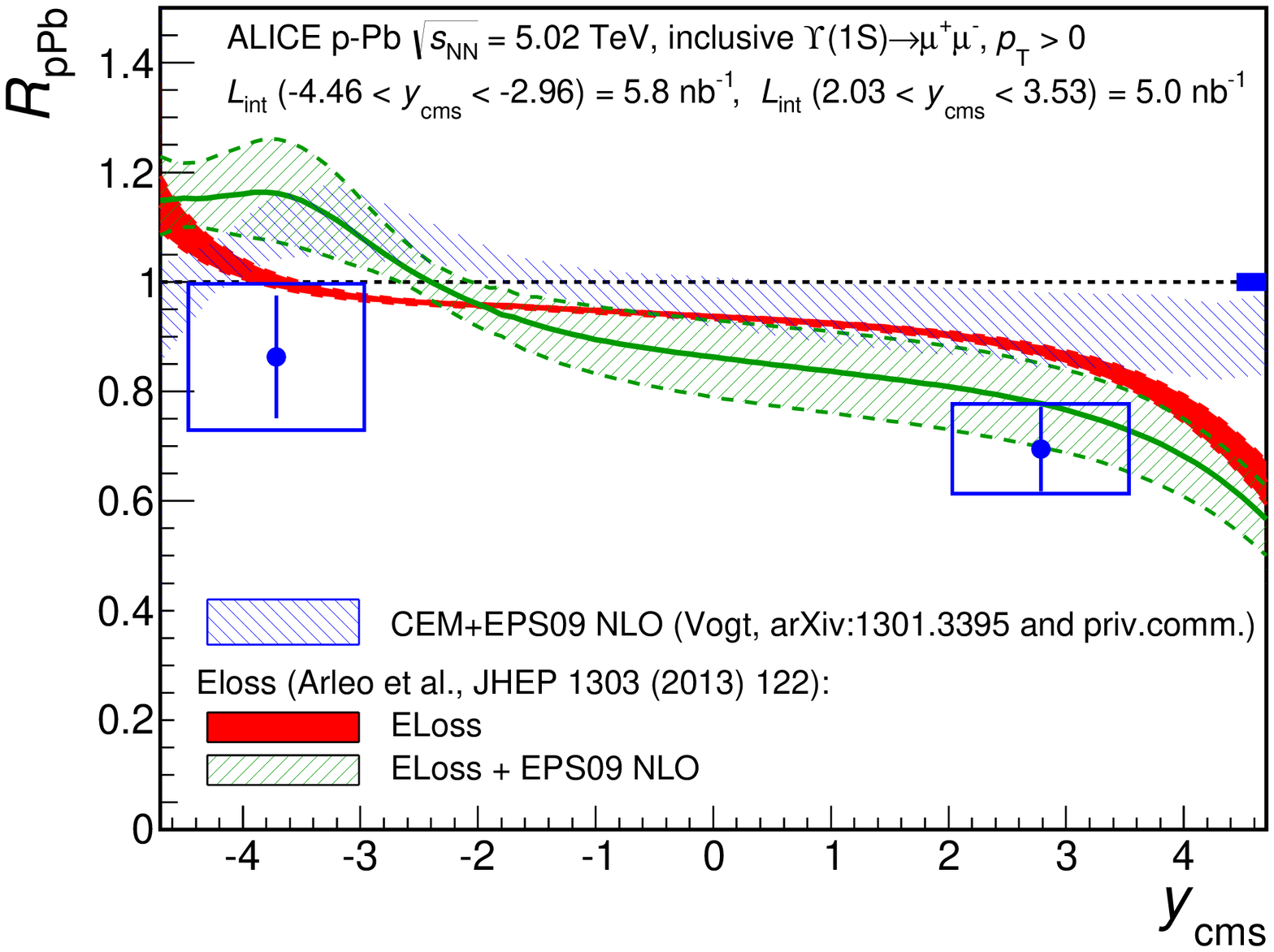} 
\includegraphics[width=0.495\linewidth]{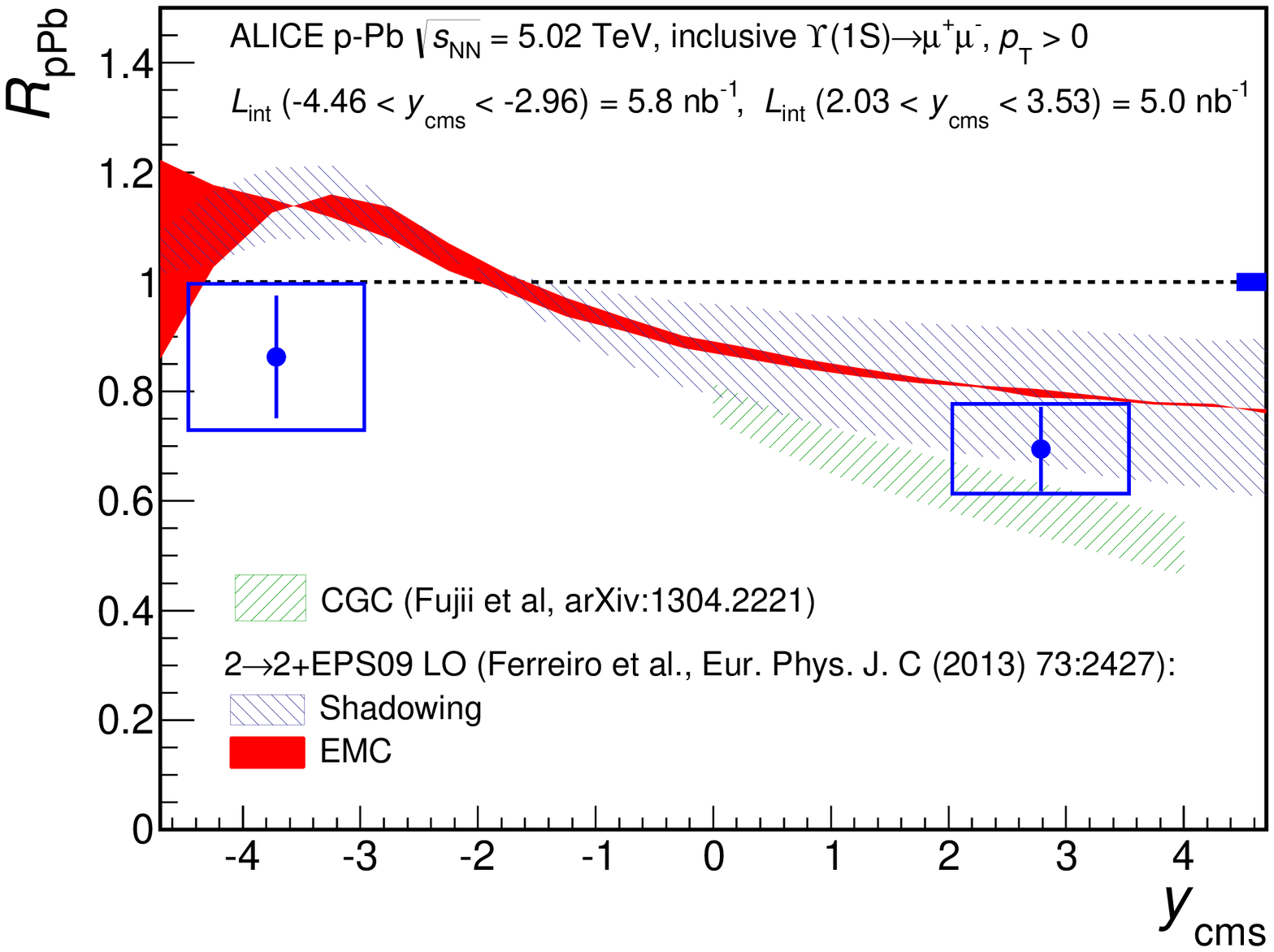} 
\end{center}
\caption{\label{fig:rppb} Nuclear modification factor of inclusive \upsis in \ppb collisions at \snn~=~5.02 TeV as a function of rapidity. The vertical error bars represent the statistical uncertainties and the open boxes the uncorrelated systematic uncertainties. The full boxes around \rpa~=~1 show the size of the correlated uncertainties. Also shown are several model calculations: (left) parton energy loss~\cite{Arleo:2012rs} with and without EPS09 shadowing at NLO and CEM with EPS09 shadowing at NLO~\cite{Albacete:2013ei}; (right) CGC based~\cite{Fujii:2013gxa} and CSM with EPS09 shadowing at LO~\cite{Ferreiro:2011xy}. For the latter the effect of variation in the shadowing and EMC curves is highlighted as described in the text.}
\end{figure}

The inclusive \upsis \rpa\ integrated over the backward or forward rapidity ranges, are compared to several model calculations in Fig.~\ref{fig:rppb}. 
In the left panel, the results are compared to a next-to-leading order (NLO) CEM calculation using the EPS09 parameterization of the nuclear modification of the gluon PDF (commonly referred to as gluon shadowing) at NLO~\cite{Albacete:2013ei} (blue shaded band) and to a parton energy loss calculation~\cite{Arleo:2012rs} with (green shaded band) or without (red band) EPS09 gluon shadowing at NLO.  
In the case of the CEM+EPS09 calculation, the band reflects the uncertainties of the calculation, dominated by the ones of the EPS09 parameterization~\cite{Eskola:2009uj}. 
In the cases of the parton energy loss model calculations, the bands represent the uncertainty from the EPS09 parameterization or from the parton transport coefficient and the parameterization used for the pp reference cross section.
None of the calculations fully describe the backward and forward rapidity data and all tend to overestimate the observed \upsis \rpa.
The parton energy loss with EPS09 calculation reproduces the \upsis \rpa\ at forward rapidity but tend to overestimate it at backward rapidity. The opposite trend is found if only parton energy loss is considered.

In the right panel, the results are compared to a calculation of a $2 \rightarrow 2$ production model ($gg \rightarrow {\Upsilon}g$) at leading order (LO) using the EPS09 shadowing parameterization also at LO~\cite{Ferreiro:2011xy}. 
Two bands are shown to highlight the uncertainties linked to two different effects. 
The extent of the blue band shows the EPS09 LO related uncertainties in the shadowing region, \emph{i.e.} at low $x_{\rm Bj}$. 
The red band shows the uncertainty in the EMC region, \emph{i.e.} at high $x_{\rm Bj}$. 
As the authors of~\cite{Ferreiro:2011xy} discuss, the gluon nPDF is poorly known in this region and the \upsis \rpa\ at backward rapidity could add useful constraints to the model calculations. 
It is worth noting that the two blue bands in the left and right panels of Fig.~\ref{fig:rppb} differ by their central curve and the extent of the uncertainties. 
The two approaches are similar and although the production models used are different, most of the difference comes from the usage of the NLO or LO EPS09 gluon shadowing parameterizations. 
It can be argued that using an NLO parameterization is more appropriate than an LO one, however it is worth remarking that other gluon shadowing parameterizations~\cite{deFlorian:2003qf,deFlorian:2011fp} (also at NLO) are available and that the uncertainty band of the EPS09 LO parameterization practically includes them. 
Therefore, the blue uncertainty band in the right panel of Fig.~\ref{fig:rppb} can be considered as including the uncertainty due to different gluon shadowing parameterizations. 
The backward rapidity \upsis \rpa\ disfavours the strong gluon anti-shadowing included in the EPS09 parameterization.
In the right panel of Fig.~\ref{fig:rppb}, a calculation based on the CGC framework coupled with a CEM production model is also shown (green shaded band) for positive $y_{{\rm cms}}$. 
It is worth noting that this calculation, although only slightly underestimating the \upsis \rpa, is not able to reproduce the \jpsi \rpa\ in the same rapidity range~\cite{Abelev:2013yxa}.

\begin{figure*}
\begin{center}
\includegraphics[width=0.495\linewidth]{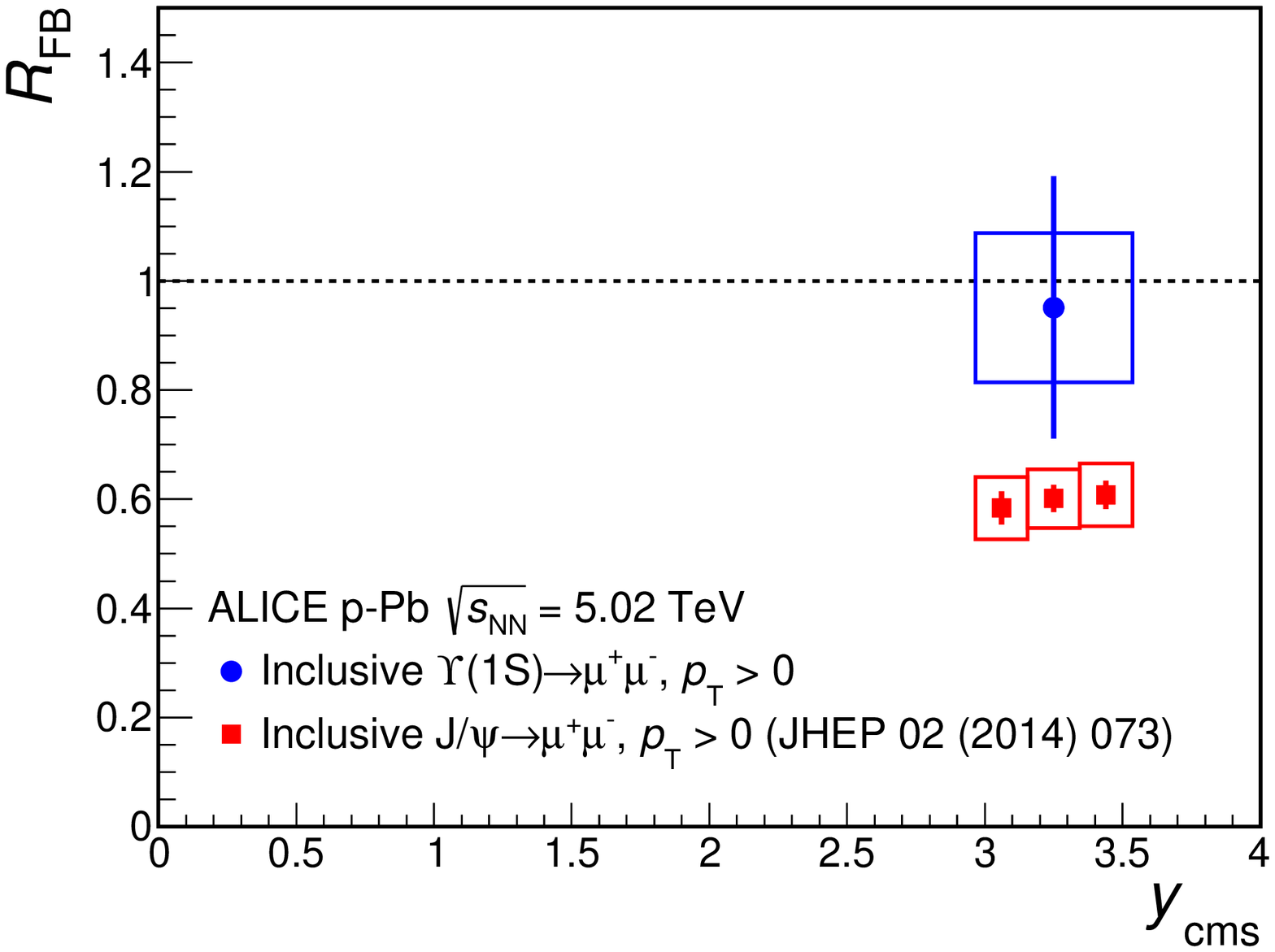} 
\includegraphics[width=0.495\linewidth]{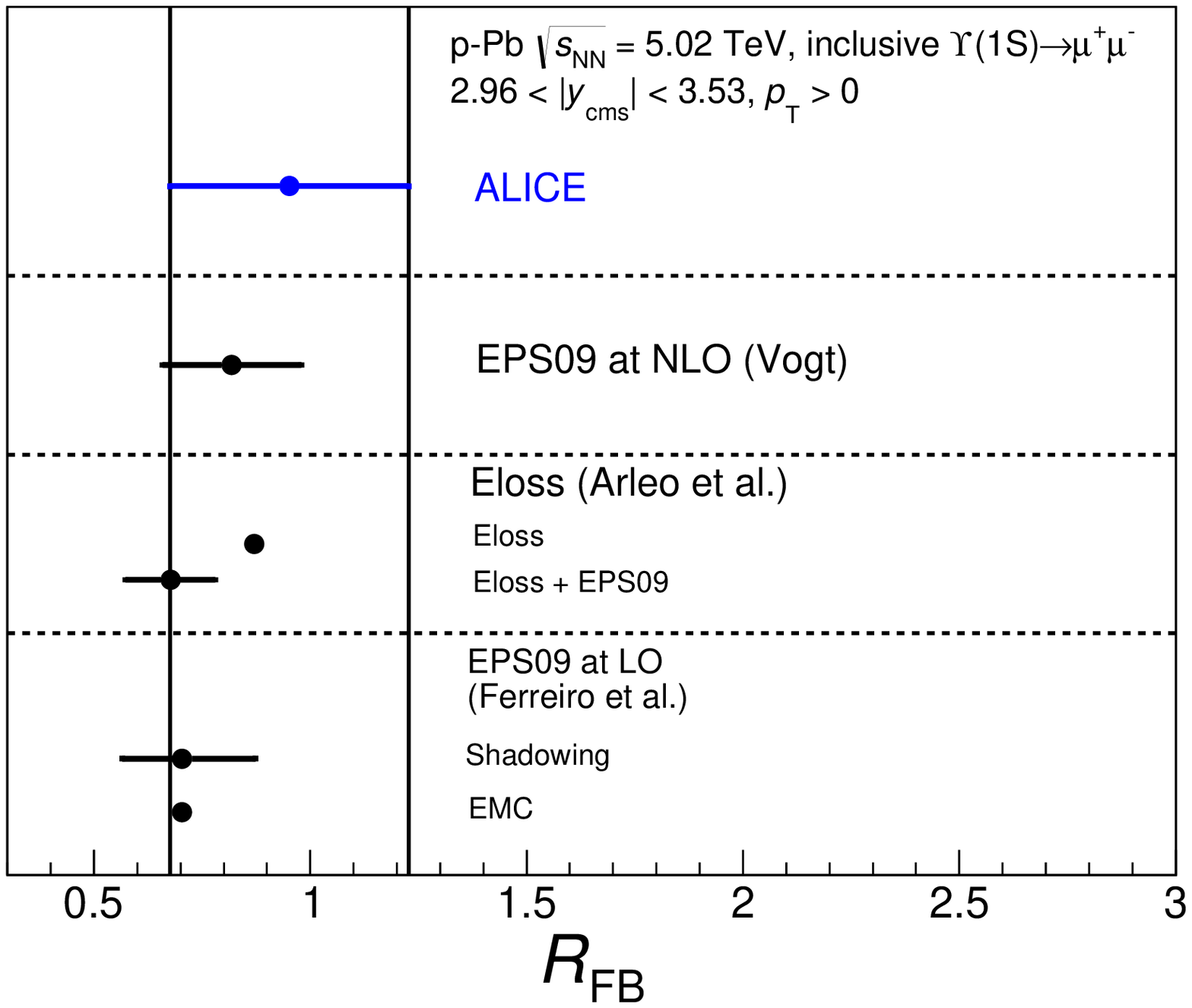} 
\end{center}
\caption{\label{fig:rfb} (Left) Forward to backward ratio \rfb\ of inclusive \upsis yields compared to the \jpsi \rfb~\cite{Abelev:2013yxa}. The vertical error bars represent the statistical uncertainties and the open boxes the uncorrelated systematic uncertainties. (Right) Inclusive \upsis \rfb\ compared to theoretical model calculations. The statistical and systematic uncertainties for the experimental value are added in quadrature. For the calculations, uncertainties are quoted when available.} 
\end{figure*}

The quantity \rfb\ is defined as the ratio of the nuclear modification factors at forward and at backward rapidities in a range symmetric with respect to $y_{{\rm cms}}=0$. 
It can be computed directly from the ratio of the cross sections (see Eq.~\ref{XSectionFormula}) of \upsis at forward and backward rapidities. 
\rfb\ is therefore independent of $\sigma^{\Upsilon\rm{(1S)}}_{pp}$.
The drawback of the \rfb\ ratio is that it can only be measured in the restricted rapidity range $2.96 < |y_{\rm{cms}}| < 3.53$, hence losing about two thirds of the number of measured $\Upsilon$. 
The measured forward to backward ratio is $R_{\rm{FB}}(2.96 < |y_{\rm{cms}}| < 3.53) = 0.95 \pm 0.24 ({\rm stat}) \pm 0.14 ({\rm syst})$. Uncertainties are obtained by summing in quadrature the contribution of each individual element entering the ratio.
The inclusive \upsis \rfb\ is compared in Fig.~\ref{fig:rfb} to the inclusive \jpsi \rfb~\cite{Abelev:2013yxa} in the same rapidity range (left panel) and to several model calculations (right panel).
In the rapidity range $2.96 < |y_{\rm{cms}}| < 3.53$ the \upsis \rfb\ is compatible with unity and is larger than that of the J/$\psi$.
All models describe the data within the present uncertainties of the measurement.

%%% Conclusion
\section{Conclusion}
In summary, we reported the ALICE measurement of \upsi production in \ppb collisions at \snn~=~5.02 TeV at the LHC. 
The \upsis production cross section and nuclear modification factor were presented in the rapidity ranges $-4.46 < y_{\rm cms} < -2.96$ and $2.03 < y_{\rm cms} < 3.53$ down to zero transverse momentum. 
At forward rapidity, \rpa\ shows a suppression of \upsis production in \ppb compared to \pp collisions. 
At backward rapidity, the \upsis \rpa\ is consistent with unity, suggesting that gluon anti-shadowing is smaller than expected in the EPS09 parameterization. 
Models including the nuclear modification of the gluon PDF~\cite{Ferreiro:2011xy,Albacete:2013ei} or a contribution from coherent parton energy loss~\cite{Arleo:2012rs} tend to overestimate our measured \rpa\ and cannot simultaneously describe the forward and backward rapidity suppressions. 
A CGC based model~\cite{Fujii:2013gxa} is in agreement with our \upsi results at forward rapidity but cannot describe the \jpsi \rpa~\cite{Abelev:2013yxa}. 
The forward to backward ratio \rfb\ of the inclusive \upsis yields in $2.96 < |y_{\rm cms}| < 3.53$ is compatible with unity within large uncertainties. 
Within our uncertainties, the $[\Upsilon{\rm (2S)}/\Upsilon{\rm (1S)}]$ ratio shows no evidence of different CNM effects on the two states. 
Additional measurements with higher statistics are needed to further constrain the models and extrapolate the CNM effects to \pbpb collisions.
              %%%%%%%%%%% put the body of the article here
%
%

%%%%% acknowledgements
\newenvironment{acknowledgement}{\relax}{\relax}
\begin{acknowledgement}
\section*{Acknowledgements}
The ALICE Collaboration would like to thank all its engineers and technicians for their invaluable contributions to the construction of the experiment and the CERN accelerator teams for the outstanding performance of the LHC complex.
%\\
The ALICE Collaboration gratefully acknowledges the resources and support provided by all Grid centres and the Worldwide LHC Computing Grid (WLCG) collaboration.
%\\
The ALICE Collaboration acknowledges the following funding agencies for their support in building and
running the ALICE detector:
 %\\
State Committee of Science,  World Federation of Scientists (WFS)
and Swiss Fonds Kidagan, Armenia,
 %\\
Conselho Nacional de Desenvolvimento Cient\'{\i}fico e Tecnol\'{o}gico (CNPq), Financiadora de Estudos e Projetos (FINEP),
Funda\c{c}\~{a}o de Amparo \`{a} Pesquisa do Estado de S\~{a}o Paulo (FAPESP);
 %\\
National Natural Science Foundation of China (NSFC), the Chinese Ministry of Education (CMOE)
and the Ministry of Science and Technology of China (MSTC);
 %\\
Ministry of Education and Youth of the Czech Republic;
 %\\
Danish Natural Science Research Council, the Carlsberg Foundation and the Danish National Research Foundation;
 %\\
The European Research Council under the European Community's Seventh Framework Programme;
 %\\
Helsinki Institute of Physics and the Academy of Finland;
 %\\
French CNRS-IN2P3, the `Region Pays de Loire', `Region Alsace', `Region Auvergne' and CEA, France;
 %\\
German BMBF and the Helmholtz Association;
%\\
General Secretariat for Research and Technology, Ministry of
Development, Greece;
%\\
Hungarian OTKA and National Office for Research and Technology (NKTH);
 %\\
Department of Atomic Energy and Department of Science and Technology of the Government of India;
 %\\
Istituto Nazionale di Fisica Nucleare (INFN) and Centro Fermi -
Museo Storico della Fisica e Centro Studi e Ricerche "Enrico
Fermi", Italy;
 %\\
MEXT Grant-in-Aid for Specially Promoted Research, Ja\-pan;
 %\\
Joint Institute for Nuclear Research, Dubna;
 %\\
%Korea Foundation for International Cooperation of Science and Technology (KICOS);
National Research Foundation of Korea (NRF);
 %\\
CONACYT, DGAPA, M\'{e}xico, ALFA-EC and the EPLANET Program
(European Particle Physics Latin American Network)
 %\\
Stichting voor Fundamenteel Onderzoek der Materie (FOM) and the Nederlandse Organisatie voor Wetenschappelijk Onderzoek (NWO), Netherlands;
 %\\
Research Council of Norway (NFR);
 %\\
Polish Ministry of Science and Higher Education;
 %\\
National Science Centre, Poland;
 %\\
 Ministry of National Education/Institute for Atomic Physics and CNCS-UEFISCDI - Romania;
 %\\
Ministry of Education and Science of Russian Federation, Russian
Academy of Sciences, Russian Federal Agency of Atomic Energy,
Russian Federal Agency for Science and Innovations and The Russian
Foundation for Basic Research;
 %\\
Ministry of Education of Slovakia;
 %\\
Department of Science and Technology, South Africa;
 %\\
CIEMAT, EELA, Ministerio de Econom\'{i}a y Competitividad (MINECO) of Spain, Xunta de Galicia (Conseller\'{\i}a de Educaci\'{o}n),
CEA\-DEN, Cubaenerg\'{\i}a, Cuba, and IAEA (International Atomic Energy Agency);
 %\\
Swedish Research Council (VR) and Knut $\&$ Alice Wallenberg
Foundation (KAW);
 %\\
Ukraine Ministry of Education and Science;
 %\\
United Kingdom Science and Technology Facilities Council (STFC);
 %\\
The United States Department of Energy, the United States National
Science Foundation, the State of Texas, and the State of Ohio;
%\\
Ministry of Science, Education and Sports of Croatia and  Unity through Knowledge Fund, Croatia.

     %%%%%%% done by webmaster team
\end{acknowledgement}

%%%%%%%% Bibliography (In case of using bibtex generate the bbl requested by arXiv)
%\bibliographystyle{style}   % Put here the style file name for the paper, i.e.apsrev4-1, utphys
%\bibliography{biblio}
%\input {bibliography.tex}  
\bibliographystyle{utphys} 
\bibliography{UpsipPb13_CernPreprint}

%%%%%%%%% appendix with author list
\newpage
\appendix
\section{The ALICE Collaboration}
\label{app:collab}

% Collaboration: CERN-LHC-ALICE
% Generation Date is 2014/Jul/01

% How to use:
%%%%%%%%% appendix with author list
%\appendix
%\section{The ALICE Collaboration}
%\label{app:collab}
%\input{authors-list.tex}  %%%%%%% get the latest version before submitting

\begingroup
\small
\begin{flushleft}
B.~Abelev\Irefn{org72}\And
J.~Adam\Irefn{org38}\And
D.~Adamov\'{a}\Irefn{org80}\And
M.M.~Aggarwal\Irefn{org84}\And
G.~Aglieri~Rinella\Irefn{org35}\And
M.~Agnello\Irefn{org108}\textsuperscript{,}\Irefn{org91}\And
A.~Agostinelli\Irefn{org27}\And
N.~Agrawal\Irefn{org45}\And
Z.~Ahammed\Irefn{org128}\And
N.~Ahmad\Irefn{org18}
% ...
\renewcommand\labelenumi{\textsuperscript{\theenumi}~}

\section*{Affiliation notes}
\renewcommand\theenumi{\roman{enumi}}
\begin{Authlist}
\item \Adef{0}Deceased
\item \Adef{idp1109600}{Also at: St. Petersburg State Polytechnical University}
\item \Adef{idp3024432}{Also at: Department of Applied Physics, Aligarh Muslim University, Aligarh, India}
\item \Adef{idp3697328}{Also at: M.V. Lomonosov Moscow State University, D.V. Skobeltsyn Institute of Nuclear Physics, Moscow, Russia}
\item \Adef{idp3947968}{Also at: University of Belgrade, Faculty of Physics and "Vin\v{c}a" Institute of Nuclear Sciences, Belgrade, Serbia}
\item \Adef{idp4280640}{Permanent Address: Konkuk University, Seoul, Korea}
\item \Adef{idp4844352}{Also at: Institute of Theoretical Physics, University of Wroclaw, Wroclaw, Poland}
\item \Adef{idp5771040}{Also at: University of Kansas, Lawrence, KS, United States}
\end{Authlist}

\section*{Collaboration Institutes}
\renewcommand\theenumi{\arabic{enumi}~}
\begin{Authlist}

\item \Idef{org1}A.I. Alikhanyan National Science Laboratory (Yerevan Physics Institute) Foundation, Yerevan, Armenia
\item \Idef{org2}Benem\'{e}rita Universidad Aut\'{o}noma de Puebla, Puebla, Mexico
\item \Idef{org3}Bogolyubov Institute for Theoretical Physics, Kiev, Ukraine
\item \Idef{org4}Bose Institute, Department of Physics and Centre for Astroparticle Physics and Space Science (CAPSS), Kolkata, India
\item \Idef{org5}Budker Institute for Nuclear Physics, Novosibirsk, Russia
\item \Idef{org6}California Polytechnic State University, San Luis Obispo, CA, United States
\item \Idef{org7}Central China Normal University, Wuhan, China
\item \Idef{org8}Centre de Calcul de l'IN2P3, Villeurbanne, France
\item \Idef{org9}Centro de Aplicaciones Tecnol\'{o}gicas y Desarrollo Nuclear (CEADEN), Havana, Cuba
\item \Idef{org10}Centro de Investigaciones Energ\'{e}ticas Medioambientales y Tecnol\'{o}gicas (CIEMAT), Madrid, Spain
\item \Idef{org11}Centro de Investigaci\'{o}n y de Estudios Avanzados (CINVESTAV), Mexico City and M\'{e}rida, Mexico
\item \Idef{org12}Centro Fermi - Museo Storico della Fisica e Centro Studi e Ricerche ``Enrico Fermi'', Rome, Italy
\item \Idef{org13}Chicago State University, Chicago, USA
\item \Idef{org14}Commissariat \`{a} l'Energie Atomique, IRFU, Saclay, France
\item \Idef{org15}COMSATS Institute of Information Technology (CIIT), Islamabad, Pakistan
\item \Idef{org16}Departamento de F\'{\i}sica de Part\'{\i}culas and IGFAE, Universidad de Santiago de Compostela, Santiago de Compostela, Spain
\item \Idef{org17}Department of Physics and Technology, University of Bergen, Bergen, Norway
\item \Idef{org18}Department of Physics, Aligarh Muslim University, Aligarh, India
\item \Idef{org19}Department of Physics, Ohio State University, Columbus, OH, United States
\item \Idef{org20}Department of Physics, Sejong University, Seoul, South Korea
\item \Idef{org21}Department of Physics, University of Oslo, Oslo, Norway
\item \Idef{org22}Dipartimento di Elettrotecnica ed Elettronica del Politecnico, Bari, Italy
\item \Idef{org23}Dipartimento di Fisica dell'Universit\`{a} 'La Sapienza' and Sezione INFN Rome, Italy
\item \Idef{org24}Dipartimento di Fisica dell'Universit\`{a} and Sezione INFN, Cagliari, Italy
\item \Idef{org25}Dipartimento di Fisica dell'Universit\`{a} and Sezione INFN, Trieste, Italy
\item \Idef{org26}Dipartimento di Fisica dell'Universit\`{a} and Sezione INFN, Turin, Italy
\item \Idef{org27}Dipartimento di Fisica e Astronomia dell'Universit\`{a} and Sezione INFN, Bologna, Italy
\item \Idef{org28}Dipartimento di Fisica e Astronomia dell'Universit\`{a} and Sezione INFN, Catania, Italy
\item \Idef{org29}Dipartimento di Fisica e Astronomia dell'Universit\`{a} and Sezione INFN, Padova, Italy
\item \Idef{org30}Dipartimento di Fisica `E.R.~Caianiello' dell'Universit\`{a} and Gruppo Collegato INFN, Salerno, Italy
\item \Idef{org31}Dipartimento di Scienze e Innovazione Tecnologica dell'Universit\`{a} del  Piemonte Orientale and Gruppo Collegato INFN, Alessandria, Italy
\item \Idef{org32}Dipartimento Interateneo di Fisica `M.~Merlin' and Sezione INFN, Bari, Italy
\item \Idef{org33}Division of Experimental High Energy Physics, University of Lund, Lund, Sweden
\item \Idef{org34}Eberhard Karls Universit\"{a}t T\"{u}bingen, T\"{u}bingen, Germany
\item \Idef{org35}European Organization for Nuclear Research (CERN), Geneva, Switzerland
\item \Idef{org36}Faculty of Engineering, Bergen University College, Bergen, Norway
\item \Idef{org37}Faculty of Mathematics, Physics and Informatics, Comenius University, Bratislava, Slovakia
\item \Idef{org38}Faculty of Nuclear Sciences and Physical Engineering, Czech Technical University in Prague, Prague, Czech Republic
\item \Idef{org39}Faculty of Science, P.J.~\v{S}af\'{a}rik University, Ko\v{s}ice, Slovakia
\item \Idef{org40}Frankfurt Institute for Advanced Studies, Johann Wolfgang Goethe-Universit\"{a}t Frankfurt, Frankfurt, Germany
\item \Idef{org41}Gangneung-Wonju National University, Gangneung, South Korea
\item \Idef{org42}Gauhati University, Department of Physics, Guwahati, India
\item \Idef{org43}Helsinki Institute of Physics (HIP), Helsinki, Finland
\item \Idef{org44}Hiroshima University, Hiroshima, Japan
\item \Idef{org45}Indian Institute of Technology Bombay (IIT), Mumbai, India
\item \Idef{org46}Indian Institute of Technology Indore, Indore (IITI), India
\item \Idef{org47}Inha University, Incheon, South Korea
\item \Idef{org48}Institut de Physique Nucl\'eaire d'Orsay (IPNO), Universit\'e Paris-Sud, CNRS-IN2P3, Orsay, France
\item \Idef{org49}Institut f\"{u}r Informatik, Johann Wolfgang Goethe-Universit\"{a}t Frankfurt, Frankfurt, Germany
\item \Idef{org50}Institut f\"{u}r Kernphysik, Johann Wolfgang Goethe-Universit\"{a}t Frankfurt, Frankfurt, Germany
\item \Idef{org51}Institut f\"{u}r Kernphysik, Westf\"{a}lische Wilhelms-Universit\"{a}t M\"{u}nster, M\"{u}nster, Germany
\item \Idef{org52}Institut Pluridisciplinaire Hubert Curien (IPHC), Universit\'{e} de Strasbourg, CNRS-IN2P3, Strasbourg, France
\item \Idef{org53}Institute for Nuclear Research, Academy of Sciences, Moscow, Russia
\item \Idef{org54}Institute for Subatomic Physics of Utrecht University, Utrecht, Netherlands
\item \Idef{org55}Institute for Theoretical and Experimental Physics, Moscow, Russia
\item \Idef{org56}Institute of Experimental Physics, Slovak Academy of Sciences, Ko\v{s}ice, Slovakia
\item \Idef{org57}Institute of Physics, Academy of Sciences of the Czech Republic, Prague, Czech Republic
\item \Idef{org58}Institute of Physics, Bhubaneswar, India
\item \Idef{org59}Institute of Space Science (ISS), Bucharest, Romania
\item \Idef{org60}Instituto de Ciencias Nucleares, Universidad Nacional Aut\'{o}noma de M\'{e}xico, Mexico City, Mexico
\item \Idef{org61}Instituto de F\'{\i}sica, Universidad Nacional Aut\'{o}noma de M\'{e}xico, Mexico City, Mexico
\item \Idef{org62}iThemba LABS, National Research Foundation, Somerset West, South Africa
\item \Idef{org63}Joint Institute for Nuclear Research (JINR), Dubna, Russia
\item \Idef{org64}Konkuk University, Seoul, South Korea
\item \Idef{org65}Korea Institute of Science and Technology Information, Daejeon, South Korea
\item \Idef{org66}KTO Karatay University, Konya, Turkey
\item \Idef{org67}Laboratoire de Physique Corpusculaire (LPC), Clermont Universit\'{e}, Universit\'{e} Blaise Pascal, CNRS--IN2P3, Clermont-Ferrand, France
\item \Idef{org68}Laboratoire de Physique Subatomique et de Cosmologie, Universit\'{e} Grenoble-Alpes, CNRS-IN2P3, Grenoble, France
\item \Idef{org69}Laboratori Nazionali di Frascati, INFN, Frascati, Italy
\item \Idef{org70}Laboratori Nazionali di Legnaro, INFN, Legnaro, Italy
\item \Idef{org71}Lawrence Berkeley National Laboratory, Berkeley, CA, United States
\item \Idef{org72}Lawrence Livermore National Laboratory, Livermore, CA, United States
\item \Idef{org73}Moscow Engineering Physics Institute, Moscow, Russia
\item \Idef{org74}National Centre for Nuclear Studies, Warsaw, Poland
\item \Idef{org75}National Institute for Physics and Nuclear Engineering, Bucharest, Romania
\item \Idef{org76}National Institute of Science Education and Research, Bhubaneswar, India
\item \Idef{org77}Niels Bohr Institute, University of Copenhagen, Copenhagen, Denmark
\item \Idef{org78}Nikhef, National Institute for Subatomic Physics, Amsterdam, Netherlands
\item \Idef{org79}Nuclear Physics Group, STFC Daresbury Laboratory, Daresbury, United Kingdom
\item \Idef{org80}Nuclear Physics Institute, Academy of Sciences of the Czech Republic, \v{R}e\v{z} u Prahy, Czech Republic
\item \Idef{org81}Oak Ridge National Laboratory, Oak Ridge, TN, United States
\item \Idef{org82}Petersburg Nuclear Physics Institute, Gatchina, Russia
\item \Idef{org83}Physics Department, Creighton University, Omaha, NE, United States
\item \Idef{org84}Physics Department, Panjab University, Chandigarh, India
\item \Idef{org85}Physics Department, University of Athens, Athens, Greece
\item \Idef{org86}Physics Department, University of Cape Town, Cape Town, South Africa
\item \Idef{org87}Physics Department, University of Jammu, Jammu, India
\item \Idef{org88}Physics Department, University of Rajasthan, Jaipur, India
\item \Idef{org89}Physik Department, Technische Universit\"{a}t M\"{u}nchen, Munich, Germany
\item \Idef{org90}Physikalisches Institut, Ruprecht-Karls-Universit\"{a}t Heidelberg, Heidelberg, Germany
\item \Idef{org91}Politecnico di Torino, Turin, Italy
\item \Idef{org92}Purdue University, West Lafayette, IN, United States
\item \Idef{org93}Pusan National University, Pusan, South Korea
\item \Idef{org94}Research Division and ExtreMe Matter Institute EMMI, GSI Helmholtzzentrum f\"ur Schwerionenforschung, Darmstadt, Germany
\item \Idef{org95}Rudjer Bo\v{s}kovi\'{c} Institute, Zagreb, Croatia
\item \Idef{org96}Russian Federal Nuclear Center (VNIIEF), Sarov, Russia
\item \Idef{org97}Russian Research Centre Kurchatov Institute, Moscow, Russia
\item \Idef{org98}Saha Institute of Nuclear Physics, Kolkata, India
\item \Idef{org99}School of Physics and Astronomy, University of Birmingham, Birmingham, United Kingdom
\item \Idef{org100}Secci\'{o}n F\'{\i}sica, Departamento de Ciencias, Pontificia Universidad Cat\'{o}lica del Per\'{u}, Lima, Peru
\item \Idef{org101}Sezione INFN, Bari, Italy
\item \Idef{org102}Sezione INFN, Bologna, Italy
\item \Idef{org103}Sezione INFN, Cagliari, Italy
\item \Idef{org104}Sezione INFN, Catania, Italy
\item \Idef{org105}Sezione INFN, Padova, Italy
\item \Idef{org106}Sezione INFN, Rome, Italy
\item \Idef{org107}Sezione INFN, Trieste, Italy
\item \Idef{org108}Sezione INFN, Turin, Italy
\item \Idef{org109}SSC IHEP of NRC Kurchatov institute, Protvino, Russia
\item \Idef{org110}SUBATECH, Ecole des Mines de Nantes, Universit\'{e} de Nantes, CNRS-IN2P3, Nantes, France
\item \Idef{org111}Suranaree University of Technology, Nakhon Ratchasima, Thailand
\item \Idef{org112}Technical University of Split FESB, Split, Croatia
\item \Idef{org113}The Henryk Niewodniczanski Institute of Nuclear Physics, Polish Academy of Sciences, Cracow, Poland
\item \Idef{org114}The University of Texas at Austin, Physics Department, Austin, TX, USA
\item \Idef{org115}Universidad Aut\'{o}noma de Sinaloa, Culiac\'{a}n, Mexico
\item \Idef{org116}Universidade de S\~{a}o Paulo (USP), S\~{a}o Paulo, Brazil
\item \Idef{org117}Universidade Estadual de Campinas (UNICAMP), Campinas, Brazil
\item \Idef{org118}University of Houston, Houston, TX, United States
\item \Idef{org119}University of Jyv\"{a}skyl\"{a}, Jyv\"{a}skyl\"{a}, Finland
\item \Idef{org120}University of Liverpool, Liverpool, United Kingdom
\item \Idef{org121}University of Tennessee, Knoxville, TN, United States
\item \Idef{org122}University of the Witwatersrand, Johannesburg, South Africa
\item \Idef{org123}University of Tokyo, Tokyo, Japan
\item \Idef{org124}University of Tsukuba, Tsukuba, Japan
\item \Idef{org125}University of Zagreb, Zagreb, Croatia
\item \Idef{org126}Universit\'{e} de Lyon, Universit\'{e} Lyon 1, CNRS/IN2P3, IPN-Lyon, Villeurbanne, France
\item \Idef{org127}V.~Fock Institute for Physics, St. Petersburg State University, St. Petersburg, Russia
\item \Idef{org128}Variable Energy Cyclotron Centre, Kolkata, India
\item \Idef{org129}Vestfold University College, Tonsberg, Norway
\item \Idef{org130}Warsaw University of Technology, Warsaw, Poland
\item \Idef{org131}Wayne State University, Detroit, MI, United States
\item \Idef{org132}Wigner Research Centre for Physics, Hungarian Academy of Sciences, Budapest, Hungary
\item \Idef{org133}Yale University, New Haven, CT, United States
\item \Idef{org134}Yonsei University, Seoul, South Korea
\item \Idef{org135}Zentrum f\"{u}r Technologietransfer und Telekommunikation (ZTT), Fachhochschule Worms, Worms, Germany
\end{Authlist}
\endgroup

\end{document}